\documentclass[seceqn]{elsart}

\usepackage{graphicx}

\usepackage{amssymb}
\usepackage{mathrsfs}

\usepackage{amsmath}

\begin{document}

\begin{frontmatter}



\title{Supersymmetric nonperturbative formulation of the WZ model
in lower dimensions}


\author{Daisuke Kadoh},
\ead{kadoh@riken.jp}
\author{Hiroshi Suzuki}
\ead{hsuzuki@riken.jp}
\address{Theoretical Physics Laboratory, RIKEN, Wako 2-1, Saitama 351-0198,
Japan}

\begin{abstract}
A nonperturbative formulation of the Wess-Zumino (WZ) model in two and three
dimensions is proposed on the basis of momentum-modes truncation. The
formulation manifestly preserves full supersymmetry as well as the
translational invariance and all global symmetries, while it is shown to be
consistent with the expected locality to all orders of perturbation theory. For
the two-dimensional WZ model, a well-defined Nicolai map in the formulation
provides an interesting algorithm for Monte Carlo simulations.
\end{abstract}

\begin{keyword}
Supersymmetry
\sep Other nonperturbative techniques
\sep Field theories in dimensions other than four
\PACS 11.30.Pb \sep 11.15.Tk \sep 11.10.Kk
\end{keyword}
\end{frontmatter}

\section{Introduction}
\label{sec:1}
In this Letter, we propose a nonperturbative Euclidean formulation of a
dimensional reduction of the Wess-Zumino (WZ) model~\cite{Wess:1974tw} to three
and two dimensions.\footnote{We assume that the K\"ahler potential is the flat
one, $\Phi^\dagger\Phi$, and the superpotential is cubic in the
three-dimensional model, so that the model is ultraviolet (UV) finite.} On a
nonperturbative formulation of the WZ model, there exist many preceding
studies, mostly based on spacetime or spatial lattices~\cite{Dondi:1976tx,
Nicolai:1978vc,Nicolai:1978ic,Nicolai:1978se,Banks:1982ut,Sakai:1983dg,
Elitzur:1982vh,Cecotti:1982ad,Bartels:1982ue,Bartels:1983wm,Elitzur:1983nj,
Ranft:1983ag,Aratyn:1984vi,Nojiri:1984ys,Nojiri:1985vb,Schiller:1986zx,
Golterman:1988ta,Aoyama:1998in,Beccaria:1998vi,Bietenholz:1998qq,So:1998ya,
Catterall:2001fr,Fujikawa:2001ka,Fujikawa:2001ns,Fujikawa:2002ic,
Kikukawa:2002as,Fujikawa:2002pa,Catterall:2003wd,Beccaria:2003ba,
Catterall:2003ae,Beccaria:2004pa,Bonini:2004pm,Beccaria:2004ds,D'Adda:2004jb,
Giedt:2004qs,Kirchberg:2004vm,Kikukawa:2004dd,Bonini:2005qx,Giedt:2005ae,
Bergner:2007pu,Catterall:2007kn,Kastner:2008zc,D'Adda:2009kj,
Synatschke:2009nm}.\footnote{In~Sec.~2.2, we will clarify the relation of our
proposal to a lattice formulation in~Ref.~\cite{Bartels:1983wm} that is based
on the SLAC derivative~\cite{Drell:1976bq,Drell:1976mj}.}
(For exact renormalization group approaches to the WZ model, see
Refs.~\cite{Rosten:2008ih,Sonoda:2009df}.)
The desired features of our present proposal are:
(I)~full supersymmetry (SUSY) as well as the translational invariance and all
global symmetries are manifestly preserved, (II)~it is amenable to
nonperturbative studies by, for example, Monte Carlo simulations,
(III)~the formulation of the two-dimensional (2D) $\mathcal{N}=(2,2)$ WZ model
possesses a well-defined Nicolai map~\cite{Nicolai:1979nr,Nicolai:1980jc} (see
also Refs.~\cite{Parisi:1982ud,Cecotti:1983up}). On the other hand, the
locality and the reflection positivity are not manifest in our formulation and
we will show that there is actually no problem concerning these in
lower-dimensional models, at least to all orders of perturbation theory.
Therefore, we propose our formulation as a nonperturbative definition of the WZ
model in lower dimensions, although its nonperturbative validity still remains
to be examined by using, for example, numerical simulations.

Our idea is very simple. The off-shell super multiplets in the WZ model (the
chiral and anti-chiral multiplets) are expressed by the chiral and anti-chiral
superfields, $\Phi$ and~$\Phi^\dagger$.\footnote{We follow the notational
conventions of~Ref.~\cite{Wess:1992cp}, except that we consider Euclidean field
theory in terms of momentum modes. (The Fourier transformation is defined
by~$\Phi(x,\theta,\bar\theta)
=\frac{1}{L^d}\sum_pe^{ipx}\,\tilde\Phi(p,\theta,\bar\theta)$.) The
Euclidean time~$x_0$ is defined from the Lorentzian time~$x^0$ by~$x^0\to-ix_0$
and we set
$\sigma_0\equiv\bar\sigma_0\equiv\begin{pmatrix}-i&0\\0&-i\\\end{pmatrix}$.
Summation over repeated indices is always meant and the Greek index~$\mu$
runs from $0$ to~$d-1$, where $d\leq4$ is the spacetime dimension. Although we
consider the system with a single chiral multiplet for notational simplicity,
the generalization to cases with multi chiral multiplets is straightforward.}
In the momentum space,
\begin{align}
   &\tilde\Phi(p,\theta,\bar\theta)=e^{-\theta\sigma_\mu\bar\theta p_\mu}
   \left[\tilde A(p)
   +\sqrt{2}\theta\tilde\psi(p)+\theta\theta\tilde F(p)\right],
\nonumber\\
   &\tilde\Phi^\dagger(p,\theta,\bar\theta)=e^{\theta\sigma_\mu\bar\theta p_\mu}
   \left[\tilde A^*(p)
   +\sqrt{2}\bar\theta\tilde{\bar\psi}(p)
   +\bar\theta\bar\theta\tilde F^*(p)\right],
\label{onexone}   
\end{align}
as they satisfy the chiral constraints, $\bar D_{\dot\alpha}\tilde\Phi(p)=0$
and~$D_\alpha\tilde\Phi^\dagger(p)=0$, where covariant spinor derivatives are
given by~$D_\alpha=\partial/\partial\theta^\alpha
-\sigma_{\mu\alpha\dot\alpha}\bar\theta^{\dot\alpha}p_\mu$
and~$\bar D_{\dot\alpha}=-\partial/\partial\bar\theta^{\dot\alpha}
+\theta^\alpha\sigma_{\mu\alpha\dot\alpha}p_\mu$. On such off-shell multiplets in
the momentum space, SUSY is linearly realized and super transformations
generated by
\begin{equation}
   Q_\alpha=\frac{\partial}{\partial\theta^\alpha}
   +\sigma_{\mu\alpha\dot\alpha}\bar\theta^{\dot\alpha}p_\mu,\qquad
   \bar Q_{\dot\alpha}=-\frac{\partial}{\partial\bar\theta^{\dot\alpha}}
   -\theta^\alpha\sigma_{\mu\alpha\dot\alpha}p_\mu,
\label{onextwo}
\end{equation}
do not mix momentum modes with different momenta. This fact suggests that one
can regularize the functional integral of the model by restricting possible
momenta of off-shell super multiplets. \emph{Any\/} restriction on momenta does
not break SUSY.\footnote{It is crucial that we restrict the momentum of
off-shell super multiplets on which SUSY is linearly realized. If the super
transformations are non-linear in field variables, such as in supersymmetric
gauge theories in the Wess-Zumino gauge, super transformations mix modes with
different momenta and restriction on possible momenta breaks SUSY. In this
aspect, our formulation differs from a formulation of supersymmetric gauge
theories in~Ref.~\cite{Hanada:2007ti}, in which SUSY is expected to be exact
only in the limit that the momentum cutoff is removed. Note that the present
model has no gauge symmetry; it is clear that any restriction on possible
momenta breaks local gauge invariance.} From a perspective of the Euclidean
rotational symmetry in the infinite volume, it would be preferable to take a
rotational invariant restriction such as~$p^2\equiv p_\mu p_\mu\leq\Lambda^2$,
where $\Lambda$ is an UV cutoff. Thus, one may define a regularized partition
function of the WZ model by
\begin{equation}
   \mathcal{Z}\equiv\int\prod_{p^2\leq\Lambda^2}
   \left[
   d\tilde A(p)\,d\tilde A^*(p)\,d\tilde F(p)\,d\tilde F^*(p)
   \prod_{\alpha=1}^2d\tilde\psi_\alpha(p)
   \prod_{\dot\alpha=\dot1}^{\dot2}d\tilde{\bar\psi}_{\dot\alpha}(p)
   \right]
   e^{-S}.
\label{onexthree}
\end{equation}
To make this expression fully well-defined, we may assume that the system is
put in a Euclidean box of size~$L$ and the momentum~$p$ is discrete,
$p_\mu=(2\pi/L)n_\mu$, where $n_\mu\in\mathbb{Z}$. The Euclidean action~$S$ in
the momentum space reads
\begin{align}
   S\equiv
   -\int d^4\theta\,
   (\tilde\Phi^\dagger*\tilde\Phi)(0,\theta,\bar\theta)
   -\int d^2\theta\,W(\tilde\Phi)(0,\theta,\bar\theta)
   -\text{h.c.},
\label{onexfour}
\end{align}
where the symbol~$*$ denotes the convolution
\begin{equation}
   (\tilde\Phi_1*\tilde\Phi_2)(p,\theta,\bar\theta)
   \equiv\frac{1}{L^d}\sum_q
   \tilde\Phi_1(q,\theta,\bar\theta)\tilde\Phi_2(p-q,\theta,\bar\theta),
\label{onexfive}
\end{equation}
and the product in the superpotential~$W(\tilde\Phi)(0,\theta,\bar\theta)$
in~Eq.~(\ref{onexfour}) is defined by repeated applications of this convolution
rule. We see that action~(\ref{onexfour}) is invariant under a multiplication
of operators~(\ref{onextwo}) on each field variables, because the sum of
momenta in action~(\ref{onexfour}) is zero, corresponding to the translational
invariance.

Prescription~(\ref{onexthree}) thus manifestly preserves SUSY as well as the
translational invariance. All global symmetries of the action (such as the
$R$-symmetry) are also manifest. Specifically, one can derive Ward-Takahashi
identities associated with these symmetries within a regularized framework.
Since definition~(\ref{onexthree}) does not modify the spinor-space structure
of the WZ model, one may repeat the proof of perturbative non-renormalization
theorems~\cite{Fujikawa:1974ay,West:1976wz,Grisaru:1979wc} in this regularized
framework (cf.~Ref.~\cite{Wess:1992cp}). One may also repeat the argument
of~Ref.~\cite{Seiberg:1994bp} because the holomorphy is manifestly preserved.
Furthermore, prescription~(\ref{onexthree}) is amenable to nonperturbative
studies by, for example, Monte Carlo simulations, because Eq.~(\ref{onexthree})
is a finite-dimensional integral for finite $\Lambda$ and~$L$.

\section{Locality}
\subsection{Locality and finiteness}
The above description sounds too good to be true. Actually, it is not clear
whether definition~(\ref{onexthree}) (in the limit that the UV and IR cutoffs
are removed, $\Lambda\to\infty$ and~$L\to\infty$) is consistent with the
expected locality in the target theory. The reflection positivity is a related
issue. Prescription~(\ref{onexthree}) differs from the conventional momentum
cutoff in perturbative Feynman integrals, with which the locality would be
obvious. The point is that the restriction $p^2\leq\Lambda^2$ on integration
variables in~Eq.~(\ref{onexthree}) could introduce non-smooth dependence of a
Feynman integral on external momenta; such dependence could not be interpreted
as an insertion of local operators. See the analysis in~Appendix.\footnote{In
this subsection and in Appendix, where the issue of locality is addressed, we
set $L\to\infty$ because the notion of locality becomes transparent only in
this limit.}

Now, if the model is massive, $W(\Phi)=(1/2)m\Phi^2+\cdots$, the free super
propagators are given by
\begin{align}
   &\left\langle\tilde\Phi(p,\theta,\bar\theta)
   \tilde\Phi^\dagger(q,\theta',\bar\theta')\right\rangle
   =\frac{1}{16}\bar D^2D^2
   \frac{1}{p^2+|m|^2}(2\pi)^d\delta(p+q)
   \delta(\theta-\theta')\delta(\bar\theta-\bar\theta'),
\label{twoxone}\\
   &\left\langle\tilde\Phi^\dagger(p,\theta,\bar\theta)
   \tilde\Phi(q,\theta',\bar\theta')\right\rangle
   =\frac{1}{16}D^2\bar D^2
   \frac{1}{p^2+|m|^2}(2\pi)^d\delta(p+q)
   \delta(\theta-\theta')\delta(\bar\theta-\bar\theta'),
\label{twoxtwo}\\
   &\left\langle\tilde\Phi(p,\theta,\bar\theta)
   \tilde\Phi(q,\theta',\bar\theta')\right\rangle
   =\frac{1}{4}\bar D^2
   \frac{m^*}{p^2+|m|^2}(2\pi)^d\delta(p+q)
   \delta(\theta-\theta')\delta(\bar\theta-\bar\theta'),
\label{twoxthree}\\
   &\left\langle\tilde\Phi^\dagger(p,\theta,\bar\theta)
   \tilde\Phi^\dagger(q,\theta',\bar\theta')\right\rangle
   =\frac{1}{4}D^2
   \frac{m}{p^2+|m|^2}(2\pi)^d\delta(p+q)
   \delta(\theta-\theta')\delta(\bar\theta-\bar\theta'),
\label{twoxfour}
\end{align}
where the momentum contained in $D_\alpha$ and~$\bar D_{\dot\alpha}$ is~$p$.
In prescription~(\ref{onexthree}), the functional integral is defined in terms
of momentum modes. One may then define the field variable in the real
space by~$\Phi(x,\theta,\bar\theta)
=\int_{p^2\leq\Lambda^2}
\frac{d^dp}{(2\pi)^d}\,e^{ipx}\,\tilde\Phi(p,\theta,\bar\theta)$. Then free
super propagators in the real space are proportional to
\begin{align}
   &\int_{p^2\leq\Lambda^2}
   \frac{d^dp}{(2\pi)^d}\,\frac{e^{ip(x-y)}}{p^2+|m|^2}
\nonumber\\
   &=\frac{(2\pi)^{d/2}}{\left|x-y\right|^{d/2-1}}
   \left[|m|^{d/2-1}K_{d/2-1}\left(|m||x-y|\right)
   -\int_\Lambda^\infty dp\,
   \frac{p^{d/2}J_{d/2-1}\left(p|x-y|\right)}{p^2+|m|^2}\right],
\end{align}
where $J_\nu(z)$ ($K_\nu(z)$) denotes the (modified) Bessel function. In the
right-hand side, the first term is the standard massive propagator which dumps
exponentially~$\sim e^{-|m||x-y|}$. The second term is the cutoff effect and its
amplitude dumps only in the inverse powers of~$|x-y|$. Therefore, when
$\Lambda$ is kept fixed, the second term dominates the first for $|x-y|$ large.
Nevertheless, since the second term is integrable at~$p\to\infty$, the second
term with $|x-y|$ kept fixed vanishes as~$\Lambda\to\infty$. Free propagators
in the real space thus restore the expected locality for~$\Lambda\to\infty$.


Next, we consider the effect of interaction. In prescription~(\ref{onexthree}),
only momentum modes with $p^2\leq\Lambda^2$ appear and, in perturbation theory,
this restriction can be taken into account by substituting all factors
$1/(p^2+|m|^2)$ in~Eqs.~(\ref{twoxone})--(\ref{twoxfour})
by~$\Theta(\Lambda^2-p^2)/(p^2+|m|^2)$, where $\Theta(x)$ denotes the step
function. The locality is not obvious in general with this prescription as
illustrated in~Appendix. Nevertheless, if a convergence property of a Feynman
integral is good enough, the value of the Feynman integral must be independent
of the regularization as~$\Lambda\to\infty$; then the issue of locality and
reflection positivity should not matter. Since our formulation preserves
manifest SUSY, we expect a better convergence property of Feynman integrals
compared with formulations which do not have manifest SUSY.\footnote{If SUSY
were not manifest, even perturbation theory in a dimensional reduction of the
WZ model to one dimension would suffer from logarithmic divergences
(see~Ref.~\cite{Giedt:2004vb}).}

Since SUSY is manifest with prescription~(\ref{onexthree}), one can determine
the superficial degrees of divergence on the basis of the super Feynman rule
(see~Ref.~\cite{Wess:1992cp}). In the WZ model in $d$~dimensions, the
superficial degrees of divergence~$\omega({\mit\Gamma})$ of a super
diagram~$\mit\Gamma$ is given by (cf.~Sec.~6.6 of~Ref.~\cite{Gates:1983nr})
\begin{equation}
   \omega({\mit\Gamma})=d-2-\frac{1}{2}(d-2)E
   +\sum_i\left[\frac{1}{2}(d-2)i+1-d\right]V_i-C,
\label{twoxsix}
\end{equation}
where $E$ denotes the number of external lines, $V_i$ the number of the
$\Phi^i$-type interaction vertex and $C$ the number of the (anti-)chiral
propagators, Eqs.~(\ref{twoxthree}) and~(\ref{twoxfour}).

For the four-dimensional (4D) system, $d=4$, the perturbative renormalizability
requires the superpotential is cubic $W(\Phi)=(1/2)m\Phi^2+(1/3)g\Phi^3$. Then
Eq.~(\ref{twoxsix}) yields $\omega({\mit\Gamma})=2-E-C$ and we see that only
two-point functions may logarithmically diverge (tadpoles, for which $E=1$,
identically vanish owing to the non-renormalization theorem;
see~Ref.~\cite{Wess:1992cp}). For such divergent diagrams, the locality is not
obvious. In fact, as explained in~Appendix, we could neither prove nor disprove
consistency of prescription~(\ref{onexthree}) with the expected locality beyond
the one-loop level. Thus we must admit that the validity of
prescription~(\ref{onexthree}) is not clear for 4D WZ model.

For the three-dimensional (3D) system, $d=3$, if the superpotential is cubic
$W(\Phi)=(1/2)m\Phi^2+(1/3)g\Phi^3$, we have
$\omega({\mit\Gamma})=1-(1/2)E-(1/2)V_3-C$ and, since again the tadpoles
identically vanish owing to SUSY, all Feynman diagrams have strictly negative
superficial degrees of divergence.

For the two-dimensional (2D) system, $d=2$, for any (polynomial)
superpotential, we have $\omega({\mit\Gamma})=-\sum_iV_i-C$ and again all
Feynman diagrams have strictly negative superficial degrees of divergence.

The above counting shows that, in 3D $\mathcal{N}=2$ WZ model with the cubic
superpotential and in 2D $\mathcal{N}=(2,2)$ WZ model with arbitrary
superpotential, all 1PI diagrams have strictly negative superficial degrees of
divergence. Combined this with the power-counting
theorem~\cite{Weinberg:1959nj,Hahn}, we see that all Feynman integrals in these
lower-dimensional models are \emph{absolutely convergent}. We then intuitively
expect that, owing to this good convergence property, the correct (finite)
value of Feynman diagrams is reproduced with prescription~(\ref{onexthree}) in
the $\Lambda\to\infty$ limit. Then there will be no need to worry about the
locality and the reflection positivity for~$\Lambda\to\infty$.

This natural expectation is rigorously confirmed by the following
\begin{lem}
For any Feynman integral
\begin{equation}
   I_F(p)\equiv\int d^dk_1\cdots d^dk_L\,\mathcal{I}_F(k,p),
\label{twoxseven}
\end{equation}
where $k_i$ are loop momenta and $p$~collectively denotes external momenta,
that is absolutely convergent
$\int d^dk_1\cdots d^dk_L\,|\mathcal{I}_F(k,p)|<\infty$, we have
\begin{equation}
   \lim_{\Lambda\to\infty}\int d^dk_1\cdots d^dk_L\,\mathcal{I}_F'(k,p;\Lambda)
   =I_F(p)
\label{twoxeight}
\end{equation}
for any fixed~$p$, where $\mathcal{I}_F'(k,p;\Lambda)$ is a modified integrand
that is defined by substituting all propagators $1/(\ell_i^2+|m|^2)$
in the original integrand~$\mathcal{I}_F(k,p)$
by~$\Theta(\Lambda^2-\ell_i^2)/(\ell_i^2+|m|^2)$.
\end{lem}
\begin{pf}
From the definition of~$\mathcal{I}_F'(k,p;\Lambda)$,
\begin{align}
   &\left|\int d^dk_1\cdots d^dk_L\,\mathcal{I}_F(k,p)
   -\int d^dk_1\cdots d^dk_L\,\mathcal{I}_F'(k,p;\Lambda)\right|
\nonumber\\
   &\leq\int d^dk_1\cdots d^dk_L\,
   \left|\mathcal{I}_F(k,p)-\mathcal{I}_F'(k,p;\Lambda)\right|
\nonumber\\
   &=\int d^dk_1\cdots d^dk_L\,
   \left|\mathcal{I}_F(k,p)\right|
   \left[1-\Theta(\Lambda^2-\ell_1)\cdots\Theta(\Lambda^2-\ell_N^2)\right],
\label{twoxnine}
\end{align}
where $N$~denotes the number of propagators and propagators' momenta~$\ell_i$
are linear combinations of~$k$ and~$p$. For fixed~$p$, for sufficiently
large~$\Lambda$, there exists a region containing the origin
of~$\mathbb{R}^{Ld}$ of size~$\Lambda'$,
$B(\Lambda')\equiv\{k_{i\mu}\in\mathbb{R}^{Ld}\mid|k_{i\mu}|\leq\Lambda'/2\}$,
such that $\Theta(\Lambda^2-\ell_1)\cdots\Theta(\Lambda^2-\ell_N^2)=1$
for~$k\in B(\Lambda')$. Setting
$\mathbb{R}^{Ld}=B(\Lambda')\cup\bar B(\Lambda')$, since
$1-\Theta(\Lambda^2-\ell_1)\cdots\Theta(\Lambda^2-\ell_N^2)=0$
for~$k\in B(\Lambda')$
and~$|1-\Theta(\Lambda^2-\ell_1)\cdots\Theta(\Lambda^2-\ell_N^2)|\leq1$, we have
\begin{equation}
   \text{Eq.~(\ref{twoxnine})}\leq
   \int_{\bar B(\Lambda')}d^dk_1\cdots d^dk_L\,
   \left|\mathcal{I}_F(k,p)\right|.
\label{twoxten}
\end{equation}
We then take the $\Lambda\to\infty$ limit on the both sides
of~Eq.~(\ref{twoxten}). The most left-hand side of~(\ref{twoxnine}) becomes
$|I_F(p)
-\lim_{\Lambda\to\infty}\int d^dk_1\cdots d^dk_L\,\mathcal{I}_F'(k,p;\Lambda)|$.
In the right-hand side of~Eq.~(\ref{twoxten}), we may then take the
$\Lambda'\to\infty$ limit and this leads to $\lim_{\Lambda'\to\infty}
\int_{\bar B(\Lambda')}d^dk_1\cdots d^dk_L\,\*|\mathcal{I}_F(k,p)|=0$
because integral~(\ref{twoxseven}) is absolutely convergent. This shows
Eq.~(\ref{twoxeight}).
\end{pf}

To summarize, to all orders of perturbation theory,
definition~(\ref{onexthree}) provides a valid formulation of 3D $\mathcal{N}=2$
WZ model with the cubic superpotential and of 2D $\mathcal{N}=(2,2)$ WZ model
with arbitrary superpotential.\footnote{Power-counting shows that if we
generalize the K\"ahler potential to an arbitrary real function
$\Phi^\dagger\Phi\to K(\Phi^\dagger,\Phi)$, new logarithmic divergences may
appear and the present argument does not apply.} We thus propose to use
Eq.~(\ref{onexthree}) as a nonperturbative definition of these models. As
already noted, all symmetries in the target theory are manifest and, moreover,
the formulation is amenable to nonperturbative Monte Carlo simulations.

\subsection{Relation to a lattice formulation based on the SLAC derivative}
As noted in~Introduction, the SUSY invariance holds with any restriction on
possible momenta of super multiplets. For example, since the rotational
symmetry is in any case broken in a finite box, one may adopt a ``cubic''
restriction~$-\Lambda\leq p_\mu\leq\Lambda$ for all~$\mu$, rather than the
``spherical one'' $p^2\leq\Lambda^2$, and
\begin{equation}
   \mathcal{Z}'\equiv\int\prod_{-\Lambda\leq p_\mu\leq\Lambda}
   \left[
   d\tilde A(p)\,d\tilde A^*(p)\,d\tilde F(p)\,d\tilde F^*(p)
   \prod_{\alpha=1}^2d\tilde\psi_\alpha(p)
   \prod_{\dot\alpha=\dot1}^{\dot2}d\tilde{\bar\psi}_{\dot\alpha}(p)
   \right]
   e^{-S},
\label{twoxeleven}
\end{equation}
where the action~$S$ is again given by~Eq.~(\ref{onexfour}). This prescription
shares desired features with Eq.~(\ref{onexthree}), such as SUSY is manifest
and the locality is restored in the $\Lambda\to\infty$ limit for
lower-dimensional models (to all orders of perturbation theory; Lemma~1 can
appropriately be modified for the cubic momentum restriction above).
Prescription~(\ref{twoxeleven}) is, however, nothing but a lattice formulation
of the WZ model in~Ref.~\cite{Bartels:1983wm} that is based on the SLAC lattice
derivative~\cite{Drell:1976bq,Drell:1976mj}.\footnote{See also
Refs.~\cite{Nojiri:1984ys,Nojiri:1985vb}
and~Refs.~\cite{Kirchberg:2004vm,Bergner:2007pu,Kastner:2008zc} for related
formulations.} In fact, if one expresses the lattice formulation
in~Ref.~\cite{Bartels:1983wm} (Eqs.~(2.10) and~(2.27) there, after the Wick
rotation) in terms of momentum modes
$\Phi(x,\theta,\bar\theta)=\frac{1}{L^d}\sum_p
e^{ipx}\,\tilde\Phi(p,\theta,\bar\theta)$, where $x$~denotes the lattice point,
one ends up with~Eq.~(\ref{twoxeleven}) with the
identification~$\Lambda\equiv\pi/a$ ($a$ is the lattice spacing).

Since two prescriptions~(\ref{onexthree}) and~(\ref{twoxeleven}) should be
essentially equivalent for~$\Lambda\to\infty$, our proposal~(\ref{onexthree})
is basically equivalent to the lattice formulation
in~Ref.~\cite{Bartels:1983wm} on the basis of the SLAC derivative.\footnote{Our
description in a previous version of this Letter on the relation to
SLAC-derivative-based lattice formulations was inadequate. We would like to
thank Yoshio Kikukawa for clarifying discussions on this point.} The SLAC
derivative is not usually adopted in lattice (gauge) theory, because the
locality could be violated~\cite{Karsten:1979wh}. See also
Refs.~\cite{Dondi:1976tx,Kato:2008sp}. This is also the case with our
prescription for 4D WZ model; as discussed in~Appendix, the consistency of
prescription~(\ref{onexthree}) with locality is not clear for 4D WZ model.
However, as we have discussed so far, the prescription can be consistent with
the locality in lower-dimensional models and we can expect the same for
prescription~(\ref{twoxeleven}).

Thus, from this perspective, our contribution in the present Letter is merely
in that we gave a strong affirmative argument for the applicability of the
formulation in~Ref.~\cite{Bartels:1983wm} to 3D $\mathcal{N}=2$ and 2D
$\mathcal{N}=(2,2)$ WZ models, dimensional reductions of the original 4D
$\mathcal{N}=1$ WZ model. On the other hand, we have to say that its validity
for 4D WZ model itself is still not clear, unfortunately.

\section{2D $\mathcal{N}=(2,2)$ WZ model}
2D $\mathcal{N}=(2,2)$ WZ model is interesting in its own right, because, for
example, it provides the Landau-Ginsburg model for $\mathcal{N}=(2,2)$
superconformal field theory~\cite{Kastor:1988ef,Martinec:1988zu,Vafa:1988uu,
Greene:1988ut,Lerche:1989uy}. In the dimensional reduction from four to two
dimensions, we set $\mu=0$ and~$\mu=3$ directions unreduced. Then,
from~Eq.~(\ref{onexfour}), in terms of component fields in~Eq.~(\ref{onexone}),
we have
\begin{align}
   S&=\frac{1}{L^2}\sum_p
   \Biggl[
   4p_z\tilde A^*(-p)p_{\bar z}\tilde A(p)-\tilde F^*(-p)\tilde F(p)
\nonumber\\
   &\qquad\qquad\qquad{}
   -\tilde F^*(-p)*W^\prime(\tilde A)^*(p)
   -\tilde F(-p)*W^\prime(\tilde A)(p)
\nonumber\\
   &\qquad\qquad\qquad{}+(\tilde{\bar\psi}_{\dot1},\tilde\psi_2)(-p)
   \begin{pmatrix}
   2ip_z&W^{\prime\prime}(\tilde A)^**\\
   W^{\prime\prime}(\tilde A)*&2ip_{\bar z}\\
   \end{pmatrix}
   \begin{pmatrix}
   \tilde\psi_1\\
   \tilde{\bar\psi}_{\dot2}
   \end{pmatrix}(p)\Biggr],
\label{threexone}
\end{align}
where we have defined $p_z\equiv(1/2)(p_0-ip_3)$
and~$p_{\bar z}\equiv(1/2)(p_0+ip_3)$; the product in~$W''(\tilde A)$ has been
defined by repeated applications of convolution~(\ref{onexfive}). Since SUSY is
manifest in our formulation, we could repeat the argument in, for example,
Sec.~2 of~Ref.~\cite{Cecotti:1989jc}. For the argument there, important
fermionic symmetries are (in the notation of~Ref.~\cite{Wess:1992cp})
\begin{align}
   &\bar Q_{\dot1}\tilde\psi_1(p)=-2\sqrt{2}ip_{\bar z}\tilde A(p),
   \qquad\bar Q_{\dot1}\tilde A(p)=0,
\nonumber\\
   &\bar Q_{\dot1}\tilde F(p)=-2\sqrt{2}ip_{\bar z}\tilde\psi_2(p),
   \qquad\bar Q_{\dot1}\tilde\psi_2(p)=0,
\nonumber\\
   &\bar Q_{\dot1}\tilde A^*(p)=\sqrt{2}\tilde{\bar\psi}_{\dot1}(p),
   \qquad\bar Q_{\dot1}\tilde{\bar\psi}_{\dot1}(p)=0,
\nonumber\\
   &\bar Q_{\dot1}\tilde{\bar\psi}_{\dot2}(p)=-\sqrt{2}\tilde F^*(p),
   \qquad\bar Q_{\dot1}\tilde F^*(p)=0,
\end{align}
and
\begin{align}
   &\bar Q_{\dot2}\tilde\psi_2(p)=-2\sqrt{2}ip_z\tilde A(p),
   \qquad\bar Q_{\dot2}\tilde A(p)=0,
\nonumber\\
   &\bar Q_{\dot2}\tilde F(p)=2\sqrt{2}ip_z\tilde\psi_1(p),
   \qquad\bar Q_{\dot2}\tilde\psi_1(p)=0,
\nonumber\\
   &\bar Q_{\dot2}\tilde A^*(p)=\sqrt{2}\tilde{\bar\psi}_{\dot2}(p),
   \qquad\bar Q_{\dot2}\tilde{\bar\psi}_{\dot2}(p)=0,
\nonumber\\
   &\bar Q_{\dot2}\tilde{\bar\psi}_{\dot1}(p)=\sqrt{2}\tilde F^*(p),
   \qquad\bar Q_{\dot2}\tilde F^*(p)=0.
\end{align}
These nilpotent symmetries imply that, among correlation functions of scalar
fields~$\tilde A(p)$, only those of zero momentum modes $\tilde A(0)$ can be
nontrivial. This follows from the fact that $\tilde A(p)$ with~$p\neq0$ are
$\bar Q_{\dot1}$ or~$\bar Q_{\dot2}$ exact and $\tilde A(p)$ are closed under
$\bar Q_{\dot1}$ and~$\bar Q_{\dot2}$.\footnote{The Witten index of the present
system (with a single chiral multiplet) is~$n$, when the superpotential
$W(\phi)$ is an $(n+1)$-th order
polynomial~\cite{Witten:1982df,Cecotti:1981fu}. Thus, for~$n\geq1$, SUSY cannot
be spontaneously broken.} Moreover, since the anti-holomorphic part of the
superpotential is $\bar Q_{\dot1}$ or~$\bar Q_{\dot2}$ exact, correlation
functions of~$\tilde A(0)$ depend on parameters in the superpotential only
holomorphically (i.e., they depend on~$g$ but not on~$g^*$). It is interesting
that our prescription provides a solid basis for these arguments which assume a
supersymmetric regularization. In the context of the Landau-Ginsburg model for
the superconformal field theory, one has to consider the massless (or critical)
limit. Since in this limit perturbation theory suffers from severe infrared
divergences, it must be important to formulate the system nonperturbatively.

It is well known that 2D $\mathcal{N}=(2,2)$ WZ model possesses the Nicolai
map~\cite{Nicolai:1979nr,Nicolai:1980jc} which is a mapping that
``trivializes'' the functional integral.\footnote{For 2D $\mathcal{N}=(2,2)$ WZ
model (on a 2D cylinder), a regularized Hermitian supercharge and an associated
Hamiltonian have been constructed and the existence of their finite limit (as
the UV cutoff goes to infinity) has rigorously been
proven~\cite{Jaffe:1987du,Jaffe:1987ys}. Although on general grounds we expect
that if the theory exists the limit that the UV cutoff is removed is unique, to
show the equivalence of the present prescription to the construction
of~Refs.~\cite{Jaffe:1987du,Jaffe:1987ys} is far beyond the scope of this
Letter. We expect that the Nicolai map provides a useful clue to show such a
nonperturbative finiteness in our approach.} Action~(\ref{threexone}) can be
rewritten as
\begin{align}
   S&=\frac{1}{L^2}\sum_p
   \Biggl[-\tilde G^*(-p)\tilde G(p)-\tilde G^*(-p)\tilde N(p)
   -\tilde G(-p)\tilde N^*(p)
\nonumber\\
   &\qquad\qquad\qquad{}+(\tilde{\bar\psi}_{\dot1},\tilde\psi_2)(-p)
   \begin{pmatrix}
   2ip_z&W^{\prime\prime}(\tilde A)^**\\
   W^{\prime\prime}(\tilde A)*&2ip_{\bar z}\\
   \end{pmatrix}
   \begin{pmatrix}
   \tilde\psi_1\\
   \tilde{\bar\psi}_{\dot2}
   \end{pmatrix}
   (p)\Biggr],
\label{threexfour}
\end{align}
where $\tilde G(p)\equiv\tilde F(p)-2ip_z\tilde A(p)$
and~$\tilde G^*(p)\equiv\tilde F^*(p)-2ip_{\bar z}\tilde A^*(p)$ are shifted
auxiliary fields and the combinations $\tilde N(p)$ and~$\tilde N^*(p)$ define
the Nicolai map
\begin{align}
   &\left\{\tilde A(p),\tilde A^*(p)\right\}
\nonumber\\
   &\mapsto
   \left\{\tilde N(p)\equiv2ip_z\tilde A(p)+W^\prime(\tilde A)^*(p),
   \tilde N^*(p)\equiv2ip_{\bar z}\tilde A^*(p)+W^\prime(\tilde A)(p)\right\}.
\label{threexfive}
\end{align}
From~Eq.~(\ref{threexfour}), one sees that the fermion determinant is
precisely the Jacobian associated with change of integration
variables~(\ref{threexfive}). Thus, after the integration over the fermion
field, the bosonic integration variables become
$\{\tilde N(p),\tilde N^*(p)\}$. Moreover, the action~$S$ becomes Gaussian
in~$\{\tilde N(p),\tilde N^*(p)\}$ after the integration over the auxiliary
fields~$\tilde G(p)$ and~$\tilde G^*(p)$. In this way, the functional integral
is trivialized by map~(\ref{threexfive}).

Note that, in functional integral~(\ref{onexthree}), the momentum of auxiliary
fields $\tilde G(p)$ and~$\tilde G^*(p)$ is restricted to~$p^2\leq\Lambda^2$.
Therefore, in~Eq.~(\ref{threexfour}), only $\tilde N(p)$ and~$\tilde N^*(p)$
with~$p^2\leq\Lambda^2$ appear. That is, in prescription~(\ref{onexthree}),
Nicolai map~(\ref{threexfive}) is a mapping
from~$\{\tilde A(p),\tilde A^*(p)\}$ to~$\{\tilde N(p),\tilde N^*(p)\}$, both
are subject of identical momentum restriction~$p^2\leq\Lambda^2$. In this
sense, the Nicolai map is well-defined with prescription~(\ref{onexthree}).

The existence of the Nicolai map provides a quite interesting simulation
algorithm for the present system. See Ref.~\cite{Beccaria:1998vi} for actual
implementation of this idea in a discretized real space. One first generates a
set of Gaussian random numbers with the unit covariance; this gives a
configuration of~$\{\tilde N(p),\tilde N^*(p)\}$. Then one inverts Nicolai
map~(\ref{threexfive}) by numerical means. There may exist several inverse
images and one must in principle find all of them. This provides
configuration(s) of~$\{\tilde A(p),\tilde A^*(p)\}$. Repeating these steps, one
obtains a statistical ensemble of~$\{\tilde A(p),\tilde A^*(p)\}$. Correlation
functions of the fermion field can also be obtained from bosonic ones without
inversion (by assuming that SUSY is not spontaneously broken). A great
advantage of this algorithm is that, compared with conventional methods based
on the Markov process, there is (in principle) \emph{no\/} autocorrelation
between configurations in the obtained ensemble.\footnote{We would like thank
Martin L\"uscher for bringing our attention to this point.
In~Ref.~\cite{Luscher:2009eq}, an algorithm on the basis of a map that
(approximately) trivializes the functional integral in lattice gauge theory has
been constructed, aiming at avoiding the critical slowing down.} Another
important point to note is that the fermion determinant in the present system
is generally complex and thus conventional methods may fail owing to the sign
problem, while the algorithm based on the Nicolai map seems to be free from
this difficulty. In the near future, we hope to carry out Monte Carlo
simulations of 2D $\mathcal{N}=(2,2)$ WZ model on the basis of the Nicolai map
in the present momentum-space formulation.

\ack

We would like to thank Tetsuo Horigane, Hiroki Kawai, Yoshio Kikukawa
and~Fumihiko Sugino for enlightening and enjoyable discussions. This work was
initially motivated by Jun~Saito's talk at the YITP workshop, ``Development of
Quantum Field Theory and String Theory'' (YITP-W-09-04). The work of H.S.\ is
supported in part by a Grant-in-Aid for Scientific Research, 18540305.

\section*{Note added}
After completing this work, we became aware of a paper by Georg
Bergner~\cite{Bergner:2009vg} in which a lattice Monte Carlo simulation of
the SUSY quantum mechanics (QM)~\cite{Witten:1982df} was carried out on the
basis of a ``full supersymmetric model''. This lattice formulation is just the
supersymmetric lattice formulation of~Ref.~\cite{Bartels:1983wm} applied to
SUSY QM. Since SUSY QM is UV finite with a supersymmetric regularization, a
variant of Lemma~1 ensures the restoration of locality in this formulation to
all orders of perturbation theory.



\appendix

\section{Locality in 4D $\mathcal{N}=1$ WZ model}

Owing to Lemma~1 in the main text, UV (absolutely) convergent Feynman integrals
are reproduced in prescription~(\ref{onexthree}) with~$\Lambda\to\infty$, and
thus it suffices to consider UV diverging Feynman diagrams. In the one-loop
level, the unique UV diverging 1PI super diagram is the two-point function
of~$\tilde\Phi$ and~$\tilde\Phi^\dagger$. (The one-loop 1PI functions
of~$\tilde\Phi$ (or~$\tilde\Phi^\dagger$) alone identically vanish owing to
manifest SUSY; cf.~Ref.~\cite{Wess:1992cp}.) With the present prescription, the
contribution of this super diagram to the 1PI effective action is
\begin{equation}
   S_{\text{eff}}
   =-\int d^4\theta\,2|g|^2\int\frac{d^4p}{(2\pi)^4}\,
   \tilde\Phi^\dagger(-p,\theta,\bar\theta)
   \tilde\Phi(p,\theta,\bar\theta)I_F(p),
\end{equation}
where the one-loop Feynman integral is given by
\begin{align}
   I_F(p)&\equiv\int\frac{d^4k}{(2\pi)^4}\,
   \frac{\Theta(\Lambda^2-k^2)}{k^2+|m|^2}
   \frac{\Theta\left(\Lambda^2-(k+p)^2\right)}{(k+p)^2+|m|^2}
\label{axtwo}
\\
   &=\int\frac{d^4k}{(2\pi)^4}\,
   \Theta(\Lambda^2-k^2)\Theta\left(\Lambda^2-(k+p)^2\right)
   \frac{k^2-(k+p)^2}{(k^2+|m|^2)^2\left[(k+p)^2+|m|^2\right]}
\label{axthree}
\\
   &\qquad{}+\int\frac{d^4k}{(2\pi)^4}\,
   \Theta(\Lambda^2-k^2)\frac{1}{\left(k^2+|m|^2\right)^2}
\label{axfour}
\\
   &\qquad{}-\int\frac{d^4k}{(2\pi)^4}\,
   \Theta(\Lambda^2-k^2)\left[1-\Theta\left(\Lambda^2-(k+p)^2\right)\right]
   \frac{1}{\left(k^2+|m|^2\right)^2}.
\label{axfive}
\end{align}
In the second equality, we have re-organized the integrand to address the
locality. First, the integral in~Eq.~(\ref{axthree}) is absolutely convergent
even without the regularization
factor~$\Theta(\Lambda^2-k^2)\Theta(\Lambda^2-(k+p)^2)$ and, according
to~Lemma~1, we may simply discard the factor
$\Theta(\Lambda^2-k^2)\Theta(\Lambda^2-(k+p)^2)\to1$ in the
limit~$\Lambda\to\infty$. Therefore, in this limit,
Eq.~(\ref{axthree}) becomes nothing but the finite part of the Feynman integral
which is given by the BPHZ subtraction scheme applied to the logarithmically
divergent integral in the original un-regularized theory. Next,
Eq.~(\ref{axfour})
$\xrightarrow{\Lambda\to\infty}(1/16\pi^2)\ln[\Lambda^2/(e|m|^2)]$ is an
ultraviolet diverging part that would be obtained in the conventional
momentum-cutoff regularization. This is a part subtracted by a local
counterterm (the wave function renormalization, in the present case) in the
BPHZ renormalization.

We have thus observed that, in the $\Lambda\to\infty$ limit,
Eqs.~(\ref{axthree}) and~(\ref{axfour}) reproduce the correct finite part and a
divergent local term corresponding to the BPHZ subtraction scheme. These two
terms are thus consistent with the expected locality. On the other hand, if
Eq.~(\ref{axfive}) does survive in the $\Lambda\to\infty$ limit, the expected
locality would be violated because Eq.~(\ref{axfive}) could not be a smooth
function of the external momentum~$p$; in other words, Eq.~(\ref{axfive})
could not be interpreted by an insertion of local operators.

To estimate Eq.~(\ref{axfive}), we note that the factor
$\Theta(\Lambda^2-k^2)\left[1-\Theta\left(\Lambda^2-(k+p)^2\right)\right]$ is
non-zero only in a region which is sandwiched in between two 3-spheres with the
radius~$\Lambda$, one has its center at $k=0$ and another at~$k=-p$. The
4-volume of this region~$\mathcal{V}(p;\Lambda)$ is given by
\begin{align}
   &\mathcal{V}(p;\Lambda)
\nonumber\\
   &\equiv
   \frac{1}{48\pi^3}\left[|p|\left(\Lambda^2-\frac{p^2}{4}\right)^{3/2}
   +\frac{3}{2}\Lambda^2|p|\left(\Lambda^2-\frac{p^2}{4}\right)^{1/2}
   +3\Lambda^4\arcsin\left(\frac{|p|}{2\Lambda}\right)\right].
\label{axsix}
\end{align}
On the other hand, we have simple bounds on the factor $1/(k^2+|m|^2)^2$
in~Eq.~(\ref{axfive}), from the consideration of the integration region,
\begin{equation}
   \frac{1}{\left(\Lambda^2+|m|^2\right)^2}
   \leq\frac{1}{\left(k^2+|m|^2\right)^2}
   \leq\frac{1}{\left[\left(\Lambda-|p|\right)^2+|m|^2\right]^2}.
\label{axseven}
\end{equation}
Therefore,
\begin{equation}
   \left|\text{Eq.~(\ref{axfive})}\right|
   \leq
   \frac{\mathcal{V}(p;\Lambda)}
   {\left[\left(\Lambda-|p|\right)^2+|m|^2\right]^2}
   \xrightarrow{\Lambda\to\infty}0,
\label{axeight}
\end{equation}
\emph{when the external momentum~$p$ is kept fixed in the limit}, because
$\mathcal{V}(p;\Lambda)=O(\Lambda^3)$ in such a limit. Eq.~(\ref{axfive})
therefore vanishes in the $\Lambda\to\infty$ limit and the expected
locality is reproduced. This shows that prescription~(\ref{onexthree}) is
consistent with the locality at least in the one-loop level.

In the two-loop level, however, the situation is much worse because there exist
diagrams in which the external momentum~$p$ in~Eq.~(\ref{axtwo}) becomes a loop
momentum that can be as large as the cutoff~$\Lambda$. The most singular
two-loop contribution to the effective action turns to be
\begin{align}
   S_{\text{eff}}
   &=\int d^4\theta\,8|g|^4\int\frac{d^4q}{(2\pi)^4}\,
   \tilde\Phi^\dagger(-q,\theta,\bar\theta)
   \tilde\Phi(q,\theta,\bar\theta)
\nonumber\\
   &\qquad{}\times\int\frac{d^4p}{(2\pi)^4}
   \frac{\Theta(\Lambda^2-p^2)}{\left(p^2+|m|^2\right)^2}
   \frac{(p+q)^2\Theta(\Lambda^2-(p+q)^2)}{(p+q)^2+|m|^2}\,
   I_F(p).
\label{axnine}
\end{align}
We would be happy, if last term~(\ref{axfive}) does not contribute in the
limit~$\Lambda\to\infty$, when Eq.~(\ref{axtwo}) is substituted
in~Eq.~(\ref{axnine}). Otherwise, since the effect of~Eq.~(\ref{axfive}) could
not be interpreted as an insertion of local operators, the expected locality
would be broken. Now, we first note that
$|\text{Eq.~(\ref{axfive})}|\geq\mathcal{V}(p;\Lambda)/(\Lambda^2+|m|^2)^2$
from~(\ref{axseven}). Since $p^2\leq\Lambda^2$ in~Eq.~(\ref{axnine}),
that is, $\Lambda^2-p^2/4\geq(3/4)\Lambda^2$,
and~$\arcsin(|p|/(2\Lambda))\geq|p|/(2\Lambda)$, we have
\begin{align}
   &\int\frac{d^4p}{(2\pi)^4}
   \frac{\Theta(\Lambda^2-p^2)}{\left(p^2+|m|^2\right)^2}
   \frac{(p+q)^2\Theta(\Lambda^2-(p+q)^2)}{(p+q)^2+|m|^2}
   \left|\text{Eq.~(\ref{axfive})}\right|
\label{axten}\\
   &\geq\frac{4+3\sqrt{3}}{128\pi^3}
   \frac{\Lambda^3}{\left(\Lambda^2+|m|^2\right)^2}
   \int\frac{d^4p}{(2\pi)^4}
   \frac{\Theta(\Lambda^2-p^2)}{\left(p^2+|m|^2\right)^2}
   \frac{(p+q)^2\Theta(\Lambda^2-(p+q)^2)}{(p+q)^2+|m|^2}\,|p|
\nonumber\\
   &\xrightarrow{\Lambda\to\infty}\frac{4+3\sqrt{3}}{1024\pi^5}.
\label{axeleven}
\end{align}
The last $\Lambda\to\infty$ limit was found by setting $m=0$ and~$q=0$; this is
possible because this does not lead to the infrared divergence. The above
relation shows that the contribution of~Eq.~(\ref{axfive})
to~Eq.~(\ref{axnine}) \emph{does\/} survive even in the
limit~$\Lambda\to\infty$; the contribution of the exotic term~(\ref{axfive}) in
fact does not vanish in the two-loop level.

Nevertheless, it is still not clear whether this leads to a breakdown of
locality. \emph{If} the $\Lambda\to\infty$ limit of~Eq.~(\ref{axten}) is
\emph{constant\/} in~$q$, then the contribution could be removed by a local
counter term---a finite wave-function renormalization. On the other hand, if the
limit of~Eq.~(\ref{axten}) is a nontrivial function such
as~$\sim\ln(|q|/\Lambda)$, the contribution cannot be removed by local
counterterms and the expected locality is broken. Unfortunately, we have not
been able to find which is really the case.

\end{document}